\documentclass[12pt]{article}
\usepackage[table]{xcolor}
\usepackage{colortbl}
\definecolor{lightgray}{gray}{0.9}
\usepackage{graphicx}
\usepackage{lscape}
\usepackage{citesort}
\usepackage{amssymb}
\usepackage{appendix}
\usepackage{multirow}

\usepackage{graphicx}
\usepackage{lscape}
\usepackage{citesort}
\usepackage{amssymb}
\usepackage{appendix}
\usepackage{multirow}

\usepackage{mathrsfs}
\begin{document}
\def\qq{\langle \bar q q \rangle}
\def\uu{\langle \bar u u \rangle}
\def\dd{\langle \bar d d \rangle}
\def\sp{\langle \bar s s \rangle}
\def\GG{\langle g_s^2 G^2 \rangle}
\def\Tr{\mbox{Tr}}
\def\figt#1#2#3{
        \begin{figure}
        $\left. \right.$
        \vspace*{-2cm}
        \begin{center}
        \includegraphics[width=10cm]{#1}
        \end{center}
        \vspace*{-0.2cm}
        \caption{#3}
        \label{#2}
        \end{figure}
    }

\def\figb#1#2#3{
        \begin{figure}
        $\left. \right.$
        \vspace*{-1cm}
        \begin{center}
        \includegraphics[width=10cm]{#1}
        \end{center}
        \vspace*{-0.2cm}
        \caption{#3}
        \label{#2}
        \end{figure}
                }

\def\ds{\displaystyle}
\def\beq{\begin{equation}}
\def\eeq{\end{equation}}
\def\bea{\begin{eqnarray}}
\def\eea{\end{eqnarray}}
\def\beeq{\begin{eqnarray}}
\def\eeeq{\end{eqnarray}}
\def\ve{\vert}
\def\vel{\left|}
\def\ver{\right|}
\def\nnb{\nonumber}
\def\ga{\left(}
\def\dr{\right)}
\def\aga{\left\{}
\def\adr{\right\}}
\def\lla{\left<}
\def\rra{\right>}
\def\rar{\rightarrow}
\def\lrar{\leftrightarrow}
\def\nnb{\nonumber}
\def\la{\langle}
\def\ra{\rangle}
\def\ba{\begin{array}}
\def\ea{\end{array}}
\def\tr{\mbox{Tr}}
\def\ssp{{\Sigma^{*+}}}
\def\sso{{\Sigma^{*0}}}
\def\ssm{{\Sigma^{*-}}}
\def\xis0{{\Xi^{*0}}}
\def\xism{{\Xi^{*-}}}
\def\qs{\la \bar s s \ra}
\def\qu{\la \bar u u \ra}
\def\qd{\la \bar d d \ra}
\def\qq{\la \bar q q \ra}
\def\gGgG{\la g^2 G^2 \ra}
\def\q{\gamma_5 \not\!q}
\def\x{\gamma_5 \not\!x}
\def\g5{\gamma_5}
\def\sb{S_Q^{cf}}
\def\sd{S_d^{be}}
\def\su{S_u^{ad}}
\def\sbp{{S}_Q^{'cf}}
\def\sdp{{S}_d^{'be}}
\def\sup{{S}_u^{'ad}}
\def\ssp{{S}_s^{'??}}

\def\sig{\sigma_{\mu \nu} \gamma_5 p^\mu q^\nu}
\def\fo{f_0(\frac{s_0}{M^2})}
\def\ffi{f_1(\frac{s_0}{M^2})}
\def\fii{f_2(\frac{s_0}{M^2})}
\def\O{{\cal O}}
\def\sl{{\Sigma^0 \Lambda}}
\def\es{\!\!\! &=& \!\!\!}
\def\ap{\!\!\! &\approx& \!\!\!}
\def\md{\!\!\!\! &\mid& \!\!\!\!}
\def\ar{&+& \!\!\!}
\def\ek{&-& \!\!\!}
\def\kek{\!\!\!&-& \!\!\!}
\def\cp{&\times& \!\!\!}
\def\se{\!\!\! &\simeq& \!\!\!}
\def\eqv{&\equiv& \!\!\!}
\def\kpm{&\pm& \!\!\!}
\def\kmp{&\mp& \!\!\!}
\def\mcdot{\!\cdot\!}
\def\erar{&\rightarrow&}
\def\olra{\stackrel{\leftrightarrow}}
\def\ola{\stackrel{\leftarrow}}
\def\ora{\stackrel{\rightarrow}}

\def\simlt{\stackrel{<}{{}_\sim}}
\def\simgt{\stackrel{>}{{}_\sim}}


\title{
         {\Large
                 {\bf
                   Radiative transition of negative to positive parity nucleon
                 }
         }
      }

\author{
   K. Azizi\thanks {e-mail:kazizi@dogus.edu.tr, kazem.azizi@cern.ch},
  H. Sundu \thanks {e-mail:hayriye.sundu@kocaeli.edu.tr} \\
   \small$\ast$ Department of Physics, Do\u gu\c s University, Ac{\i}badem-Kad{\i}k\"oy, 34722
  Istanbul, Turkey \\
  \small $\dag$ Department of Physics, Kocaeli University, 41380
Izmit, Turkey}
 \date{}

\begin{titlepage}
\maketitle
\thispagestyle{empty}

\begin{abstract}
We investigate the $N^{\ast}(1535)\rightarrow N\gamma$ transition  in
the framework of light cone QCD sum rules. In particular, using
the most general form of the interpolating current for the nucleon
as well as the distribution amplitudes of the photon, we calculate
two transition form factors  responsible for this channel
and use them to evaluate the decay width and branching ratio of the
 transition under consideration. The result obtained for the branching fraction is in a good consistency with the experimental data.

\end{abstract}

~~~PACS number(s): 13.40.Gp,  13.30.Ce, 14.20.Dh, 11.55.Hx
\end{titlepage}

\section{Introduction}

The investigation of the decay properties of hadrons plays essential role in understanding their internal structures as well as the perturbative and non-perturbative aspects of QCD. The theoretical study on the decay channels of the
 negative parity baryons, especially their radiative transitions, can be
more useful in this regard as we have limited information about the  decay properties of these states, experimentally. Some experimental studies on the photo-production and electro-production
 are planned to measure the electromagnetic form factors and multipole moments of the negative parity baryons at Jefferson Laboratory \cite{Jefferson}  and Mainz
Microton facility \cite{Mainz1,Mainz2,Mainz3}. One of the main difficulty in measuring the
magnetic moments of the excited baryons  can be related to the considerably large
width that they have. The magnetic moments of these states can be measured from the
polarization observables of the decay products of their excited resonances.
 It is hoped that the new electron beam facilities would  allow to collect a large
numbers of more precise  data in the studies of the electro-excitations of the nucleon resonances.

One of the main directions
in obtaining essential information about the internal structure and natures of the negative parity baryons is
to study their electromagnetic form factors and multipole moments both theoretically and experimentally. Such studies can also provide valuable information about their geometric shape. In this accordance,
in the present study, we investigate the radiative $N^{\ast}(1535)\rightarrow N\gamma$ transition  in
the framework of light cone QCD sum rules (LCSR), where $N^{\ast}(1535)$ is the low-lying negative parity nucleon with $J^P=\frac{1}{2}^-$. In the following, we will briefly refer to this state by $N^{\ast}$.
In particular, we  calculate
the two transition form factors $f_1^T$ and $f_2^T$ defining this channel (negative parity to positive parity nucleon) separating the contributions of the positive to positive and negative to negative parity
transitions entering the physical side of correlation function.
In the calculations, we use the most general form of the interpolating currents coupled to both the positive and negative parity nucleons and try to find the working region of the general parameter entering the general
interpolating current. We also deal with real  photon and calculate the transition form factors at zero transfered momentum squared using the photon distribution amplitudes (DAs).
In the following, we shall refer to some  studies on the spectroscopic and decay properties of the negative parity baryons.
The mass and other spectroscopic properties of the negative parity nucleons have been extensively studied compared to their decays (for instance see \cite{Chung,Jido,Oka,Lee,Kondo} and references therein).
The light-cone DAs of the nucleon and negative parity nucleon resonances was studied via
Lattice QCD in \cite{V.M. Braun}.
The magnetic moments of the negative parity baryons have been studied via QCD sum rules in \cite{Aliev} and from effective Hamiltonian approach to QCD in \cite{Narodetskii}. The  $\gamma^\ast N \to N(1535)$ and
$\gamma^\ast N \to N^\ast (1520)$ transition form factors are studied in \cite{savci1,savci2} in light cone QCD sum rules using the nucleon DAs. The  electromagnetic transitions of the
octet negative parity to octet positive parity baryons  is also studied in \cite{savci3} via light cone QCD sum rules.

The layout of the paper is as follows. In next section, we derive LCSR for the transition electromagnetic form factors under consideration. The last section
 is dedicated to the numerical analysis of the form factors, calculation of the total decay width and branching fraction of the considered radiative transition. We also compare the results obtained with the existing experimental data.
\section{ LCSR for radiative transition of negative to positive parity nucleon }

The aim of this section is to give some  details of the calculations  of LCSR
for the transition form factors defining the radiative $N^{\ast}\rightarrow N\gamma$ channel. To
fulfill this aim, we consider the following
 two-point correlation function in the presence of the background photon field:
\begin{eqnarray}\label{CorrelationFunc}
\Pi=i \int d^4x~ e^{ip\cdot x}~{\langle}0| {\cal T}\left (
\eta(x)~\bar{\eta}(0)~\right)|0{\rangle}_{\gamma},
\end{eqnarray}
where $\eta$ is the interpolating current  coupled to both the negative and positive parity nucleon and ${\cal T}$ denotes the time ordering operator.
According to the methodology of QCD sum rule, the calculation of the above mentioned
correlation function is made via following two different ways in
order to construct sum rules for the transition form factors. In the first way, one calculates it in terms of the
hadronic degrees of freedom called as hadronic side. In the second
way, it is calculated in terms of photon distribution amplitudes
(DAs) with increasing twist by the help of operator product
expansion (OPE) called as OPE side.  These two sides are then
matched to obtain the LCSR  for the form
factors under consideration. We apply a double Borel transformation with respect to
the momentum squared of the initial and final  states to both
sides to suppress the contributions of the higher states and
continuum.

\subsection{Hadronic side}

 The calculation of the hadronic side of the
correlation function in Eq.~(\ref{CorrelationFunc}) requires its saturation  with complete sets
of intermediate states having the same quantum numbers as the
interpolating current coupling to both the negative and positive parity states. After performing the four-integral over $x$, we get
\begin{eqnarray} \label{physide}
\Pi^{HAD}&=&\frac{{\langle}0|\eta|N(p,s){\rangle}}{p^2-m_N^2}{\langle}N(p,s)\gamma(q)|
N(p+q,s^{\prime}){\rangle}\frac{{\langle}N(p+q,s^{\prime})|\bar{\eta}|0{\rangle}}
{(p+q)^2-m_N^2}
\nonumber \\
&+&\frac{{\langle}0|\eta|N^{\ast}(p,s){\rangle}}{p^2-m_{N^{\ast}}^2}{\langle}N^{\ast}
(p,s)\gamma(q)|
N^{\ast}(p+q,s^{\prime}){\rangle}\frac{{\langle}N^{\ast}(p+q,s^{\prime})|\bar{\eta}|0{\rangle}} {(p+q)^2-m_{N^{\ast}}^2}
\nonumber \\
&+&\frac{{\langle}0|\eta|N(p,s){\rangle}}{p^2-m_{N}^2}{\langle}N
(p,s)\gamma(q)|
N^{\ast}(p+q,s^{\prime}){\rangle}\frac{{\langle}N^{\ast}(p+q,s^{\prime})|\bar{\eta}|0{\rangle}} {(p+q)^2-m_{N^{\ast}}^2}
\nonumber \\
&+&\frac{{\langle}0|\eta|N^{\ast}(p,s){\rangle}}{p^2-m_{N^{\ast}}^2}{\langle}N^{\ast}
(p,s)\gamma(q)|
N(p+q,s^{\prime}){\rangle}\frac{{\langle}N(p+q,s^{\prime})|\bar{\eta}|0{\rangle}} {(p+q)^2-m_{N}^2}
\nonumber \\
&+&\cdots~,
\end{eqnarray}
where $\cdots$ represents the contributions coming from the higher
states and continuum. The matrix elements in the above equation can be
parameterized  in terms of  residues $\lambda_N$ and $\lambda_{N^{\ast}}$  as well as  transition form factors $f_1$, $f_2$, $f^*_1$, $f^*_2$, $f^T_1$ and $f^T_2$ at $q^2=0$ as
\begin{eqnarray}\label{matriselements}
\langle 0 \mid
 \eta\mid N(p,s)\rangle&=&\lambda_N u_N(p, s),
\nonumber \\
\langle N(p+q,s^{\prime}) \mid
 \bar{\eta}\mid
 0\rangle&=&\lambda_{N}\bar{u}_{N}(p+q, s^{\prime}),
\nonumber \\
\langle 0 \mid
 \eta\mid N^{\ast}(p,s)\rangle&=&\lambda_{N^{\ast}}\gamma_{5} u_{N^{\ast}}(p, s),
\nonumber \\
\langle N^{\ast}(p+q,s^{\prime}) \mid
 \bar{\eta}\mid
 0\rangle&=&\lambda_{N^{\ast}}\bar{u}_{N^{\ast}}(p+q, s^{\prime})\gamma_{5},
\nonumber \\
\langle N(p,s)\gamma(q) \mid
N(p+q,s^{\prime})\rangle&=&e\varepsilon^{\mu}\bar{u}_N(p,s)\Big[f_1\gamma_{\mu}
 -\frac{i\sigma_{\mu\nu}}{2m_N}q^{\nu}f_2\Big]u_N(p+q,s^{\prime}),
 \nonumber \\
\langle N^{\ast}(p,s)\gamma(q) \mid
N^{\ast}(p+q,s^{\prime})\rangle&=&e\varepsilon^{\mu}\bar{u}_{N^{\ast}}(p,s)\Big[f_1^{\ast}\gamma_{\mu}
 -\frac{i\sigma_{\mu\nu}}{2m_{N^{\ast}}}q^{\nu}f_2^{\ast}\Big]u_{N^{\ast}}(p+q,s^{\prime}),
 \nonumber \\
\langle N(p,s)\gamma(q) \mid
N^{\ast}(p+q,s^{\prime})\rangle&=&e\varepsilon^{\mu}\bar{u}_{N}(p,s)\Big[f_1^{T}\gamma_{\mu}
 -\frac{i\sigma_{\mu\nu}}{m_N+m_{N^{\ast}}}q^{\nu}f_2^{T}\Big]\gamma_5u_{N^{\ast}}(p+q,s^{\prime}),
 \nonumber \\
\langle N^{\ast}(p,s)\gamma(q) \mid
N(p+q,s^{\prime})\rangle&=&e\varepsilon^{\mu}\bar{u}_{N^{\ast}}(p,s)\Big[f_1^{T}\gamma_{\mu}
 -\frac{i\sigma_{\mu\nu}}{m_N+m_{N^{\ast}}}q^{\nu}f_2^{T}\Big]\gamma_5u_{N}(p+q,s^{\prime}),
 \nonumber \\
\end{eqnarray}
where
$u_N$ and $u_{N^{\ast}}$ are the spinors for the positive parity $N$ and negative parity
$N^{\ast}$ nucleons, respectively.  The use of Eq.~(\ref{matriselements})
into Eq.~(\ref{physide}) is followed by summing over the spins of
the $N$ and $N^{\ast}$ baryons, i.e.
\begin{eqnarray}\label{spinor}
\sum_{s}u_{N^{(*)}}(p,s)\bar{u}_{N^{(*)}}(p,s)&=& \not\!p+m_{N^{(*)}}~.
\end{eqnarray}
As a result, we have
\begin{eqnarray} \label{physide1}
\Pi^{HAD}&=&\frac{\lambda_N^2~e}{(p^2-m_N^2)[(p+q)^2-m_N^2]}(\not\!p+m_N)\Big[(f_1+f_2)\not\!\varepsilon
-\frac{p\cdot\varepsilon}{m_N}f_2\Big](\not\!p+\not\!q+m_N)\nonumber\\
&+&\frac{\lambda_{N^{\ast}}^2~e}{(p^2-m_{N^{\star}}^2)[(p+q)^2-m_{N^{\ast}}^2]}
(-\not\!p+m_{N^{\ast}})\Big[-(f_1^{\ast}+f_2^{\ast})\not\!\varepsilon
-\frac{p\cdot\varepsilon}{m_{N^{\ast}}}f_2^{\ast}\Big](-\not\!p-\not\!q+m_{N^{\ast}})\nonumber\\
&+&\frac{\lambda_N\lambda_{N^{\ast}}~e}{(p^2-m_{N}^2)[(p+q)^2-m_{N^{\ast}}^2]}
(\not\!p+m_{N})\Big[(f_1^{T}-\frac{m_{N^{\ast}}-m_N}{m_{N^{\ast}}+m_N}f_2^{T})\not\!\varepsilon
-\frac{2p\cdot\varepsilon}{m_{N^{\ast}}+m_N}f_2^{T}\Big]
\nonumber\\
&\times&(-\not\!p-\not\!q+m_{N^{\ast}})\nonumber\\
&+&\frac{\lambda_N\lambda_{N^{\ast}}~e}{(p^2-m_{N^{\ast}}^2)[(p+q)^2-m_{N}^2]}
(-\not\!p+m_{N^{\ast}})\Big[-(f_1^{T}+\frac{m_{N^{\ast}}-m_N}{m_{N^{\ast}}+m_N}f_2^{T})
\not\!\varepsilon
-\frac{2p\cdot\varepsilon}{m_{N^{\ast}}+m_N}f_2^{T}\Big]
\nonumber\\
&\times&(\not\!p+\not\!q+m_{N})\nonumber\\
&+&\cdots~.
\end{eqnarray}
Our aim in the present study is to calculate the transition form factors $f_1^{T}$ and $f_2^{T}$ defining the radiative  $N^{\ast}\rightarrow N\gamma$ transition. For this purpose,  we use the structures
$\not\!p\not\!\varepsilon\not\!q$, $\not\!p\not\!\varepsilon$, $\not\!\varepsilon\not\!q$ and $\not\!\varepsilon$ in our calculations.
The final form of the hadronic side of the correlation function is
obtained after the application of a double Borel transformation
with respect to the initial and final momenta squared, viz.
\begin{eqnarray} \label{physideLast}
\widehat{\textbf{B}}\Pi^{HAD}&=&\Pi_1^{HAD}\not\!p\not\!\varepsilon\not\!q+\Pi_2^{HAD}\not\!p\not\!\varepsilon
+\Pi_3^{HAD}\not\!\varepsilon\not\!q+\Pi_4^{HAD}\not\!\varepsilon
+\cdots~,
\end{eqnarray}
where we have kept only the structures which we use to construct
the LCSR for the form factors under consideration. The functions
$\Pi_{i=1,2,3,4}^{HAD}$ are
obtained as
\begin{eqnarray} \label{Pi1}
\Pi_1^{HAD}&=&\frac{2e\lambda_N\lambda_{N^{\ast}}(m_{N^{\ast}}-m_N)}{m_N+m_{N^{\ast}}}f_2^T
e^{-\frac{m_{N^{\ast}}^2}{M_1^2}-\frac{m_N^2}{M_2^2}}+e\lambda_N^2\mu_N
e^{-\frac{2m_N^2}{M_2^2}}
-e\lambda_{N^{\ast}}^2\mu_{N^{\ast}}e^{-\frac{2m_{N^{\ast}}^2}{M_1^2}},
\nonumber\\
\Pi_2^{HAD}&=&2e\lambda_{N}\lambda_{N^{\ast}}(m_N+m_{N^{\ast}})
f_1^Te^{-\frac{m_{N^{\ast}}^2}{M_1^2}-\frac{m_N^2}{M_2^2}},
\nonumber\\
\Pi_3^{HAD}&=&-\frac{e\lambda_N\lambda_{N^{\ast}}[f_1^T(m_{N^{\ast}}+m_N)^2+f_2^T(m_{N^{\ast}}-m_N)^2]}
{m_N+m_{N^{\ast}}}
e^{-\frac{m_{N^{\ast}}^2}{M_1^2}-\frac{m_N^2}{M_2^2}}
\nonumber\\
&+&e\lambda_N^2\mu_N m_N e^{-\frac{2m_N^2}{M_2^2}}
+e\lambda_{N^{\ast}}^2\mu_{N^{\ast}}m_{N^{\ast}}e^{-\frac{2m_{N^{\ast}}^2}{M_1^2}},
\nonumber\\
\Pi_4^{HAD}&=&-e\lambda_N\lambda_{N^{\ast}}(m_{N^{\ast}}^2-m_N^2)(f_1^T+f_2^T)
e^{-\frac{m_{N^{\ast}}^2}{M_1^2}-\frac{m_N^2}{M_2^2}},
\end{eqnarray}
where $M_1^2$ and $M_2^{2}$ are Borel mass parameters. As it is
clear from the explicit forms of the functions $\Pi_i^{HAD}$, they also include the magnetic moments
$\mu_N=f_1+f_2$ and $\mu_{N^{\ast}}=f_1^{\ast}+f_2^{\ast}$ which
are  calculated in \cite{Aliev1} and \cite{Aliev}, respectively.

\subsection{OPE Side}

The OPE side of the  correlation function is calculated in terms
of the QCD parameters and photon DAs. To this aim we need to know the explicit form of the nucleon interpolating current coupling to both the negative and positive parity state.
It is given as
\begin{eqnarray} \label{InterpolatingCurrent}
\eta=2\varepsilon^{abc}\Big\{\Big(u^{aT}Cd^{b}\Big)\gamma_5u^{c}
+\beta\Big(u^{aT}C\gamma_5d^{b}\Big)u^{c}\Big\},
\end{eqnarray}
where $a$, $b$ and $c$ are color indices and $C$ is the charge
conjugation operator. The parameter $\beta$ is a general mixing
parameter, which we shall find its working region in next section.
We insert
the aforementioned   interpolating current into the correlation function in
Eq.~(\ref{CorrelationFunc}) and use the Wick's theorem to contract out all the quark pairs. This leads to obtain the correlation function  on OPE side in terms of the light quark propagators.

The correlation function in OPE side contains three different parts:
\begin{itemize}
    \item perturbative part,
    \item mixed part at which the
photon is radiated from  short distances and at least one of the
quarks interacts with the QCD vacuum and makes a condensate,
    \item non-perturbative part at which the photon is radiated at long distances. This contribution appears as  the matrix elements $ \langle\gamma(q)\mid\bar
q(x_{1})
 \Gamma q(x_{2})\mid0\rangle$, which are expressed in terms of the photon
 DAs having distinct twists. The $\Gamma$ appearing in this matrix element refers to the full set of Dirac matrices, namely
$\Gamma_j = \Big\{ 1,~\gamma_5,~\gamma_\alpha,~i\gamma_5
\gamma_\alpha, ~\sigma_{\alpha \beta} /\sqrt{2}\Big\}$.
\end{itemize}

The perturbative  contributions where the photon perturbatively
interacts with the quarks are acquired via replacing the
corresponding free light quark propagator with
\begin{equation}\label{rep1guy}
S^{ab}_{\alpha \beta} \Rightarrow  \left\{ \int d^4 y S^{free}
(x-y) \not\!A S^{free}(y)\right\}^{ab}_{\alpha \beta},
\end{equation}
 at which the free part of the light quark propagator is written  as
\begin{eqnarray}\label{free1guy}
S^{free}_{q} &=&\frac{i\not\!x}{2\pi^{2}x^{4}}-\frac{m_{q}}{4\pi^{2}x^{2}},\nonumber\\
\end{eqnarray}
where $q$ stands for the light $u$ or $d$ quark.
%
%

To obtain the non-perturbative contributions we replace   the light quark propagator that interacts with the  photon with
\begin{equation}\label{rep2guy}
\label{rep} S^{ab}_{\alpha \beta} \rightarrow - \frac{1}{4} \bar
q^a \Gamma_j q^b ( \Gamma_j )_{\alpha \beta}~.
\end{equation}
 The remaining quark propagators are replaced  with the full
quark propagators   which include all  the perturbative and
non-perturbative parts. The full light quark propagator is given as \cite{Balitsky,Braun2}
\begin{eqnarray}\label{heavylightguy}
S_q(x) &=&  S_q^{free} (x) - \frac{m_q}{4 \pi^2 x^2} -
\frac{\langle \bar q q \rangle}{12} \left(1 - i \frac{m_q}{4}
\not\!x \right) - \frac{x^2}{192} m_0^2 \langle \bar q q \rangle
\left( 1 - i \frac{m_q}{6}\not\!x \right) \nonumber \\ &&
 - i g_s \int_0^1 du \left[\frac{\not\!x}{16 \pi^2 x^2} G_{\mu \nu} (ux) \sigma_{\mu \nu} - u x^\mu
G_{\mu \nu} (ux) \gamma^\nu \frac{i}{4 \pi^2 x^2} \right.
\nonumber
\\ && \left. - i \frac{m_q}{32 \pi^2} G_{\mu \nu} \sigma^{\mu \nu}
\left( \ln \left( \frac{-x^2 \Lambda^2}{4} \right) + 2 \gamma_E
\right) \right],
 \end{eqnarray}
  where $\Lambda$ is a scale parameter.

As it has already mentioned, in calculations of the non-perturbative contributions, there appear   matrix elements  like $ \langle\gamma(q)\mid\bar q
 \Gamma_{i}q\mid0\rangle$.
 These matrix  elements are expanded in terms of the photon DAs with increasing twists. We present the parametrization of these matrix elements in terms of the photon DAs together with the explicit forms of the photon wave functions
as well as the related parameters in the Appendix. After very lengthy calculations, we obtain the OPE side of the correlation function in momentum space in terms of the photon DAs and other QCD parameters.
The final form of the OPE side of the correlation function is obtained by applying a Borel transformation with respect to the variables
$p^2$ and $(p+q)^2$.

 Having calculated both the hadronic and OPE sides of the correlation function in Borel scheme it is the time for matching the two sides in order to obtain LCSR for the transition form factors $f_1^T$ and
$f_2^T$, defining the radiative transition under consideration. For this purpose,  we  match  the coefficients of the structures
  $\not\!p\not\!\varepsilon\not\!q$,  $\not\!p\not\!\varepsilon$, $\not\!\varepsilon\not\!q$ and $\not\!\varepsilon$
from both sides of the correlation function. To further suppress the contributions of the higher states and continuum we also perform continuum subtraction. By simultaneous solving of four equations
obtained via matching the coefficients of the above structures from both the  hadronic and OPE sides, we get the following sum rules for
the form factors under consideration:
\begin{eqnarray}\label{sumrules}
f_1^T&=&\frac{1}{2e\lambda_{N}\lambda_{N^{\ast}}(m_N+m_{N^{\ast}})}e^{\frac{m_N^2}{M_2^2}+
\frac{m_{N^{\ast}}^2}{M_1^2}}\Pi_2^{OPE}(M^2,s_0,\beta,u_0)
\nonumber \\
f_2^T&=&\frac{1}{2e\lambda_{N}\lambda_{N^{\ast}}(m_N^2-m_{N^{\ast}}^2)}
e^{\frac{m_N^2}{M_2^2}+
\frac{m_{N^{\ast}}^2}{M_1^2}}\Big[\Pi_2^{OPE}(M^2,s_0,\beta,u_0)\Big(m_{N^{\ast}}-m_N\Big)\nonumber\\
&+&2\Pi_4^{OPE}~(M^2,s_0,\beta,u_0)\Big], \nonumber\\
\end{eqnarray}
where $s_0$ is the continuum threshold and the functions $\Pi_{i}^{OPE}(M^2,s_0,\beta,u_0)$  are very
lengthy, so we do not present their explicit forms here.
In the above equation,  $u_0$ and the new Borel parameter $M^2$ are obtained in terms of the Borel parameters $M_1^2$ and $M_2^2$ as
\begin{eqnarray}
u_0&=&\frac{M_1^2}{M_1^2+M_2^2},
\nonumber \\
M^2&=&\frac{M_1^2M_2^2}{M_1^2+M_2^2}.
\end{eqnarray}
Using the relation $\frac{M_1^2}{M_2^2}\simeq
\frac{m_{N^{\ast}}^2}{m_N^2}$, we obtain $M_1^2$ and $M_2^2$ in
terms of $M^2$ as
\begin{eqnarray}
M_1^2&=&\frac{m_N^2+m_{N^{\ast}}^2}{m_N^2}M^2,
\nonumber \\
M_2^2&=&\frac{m_N^2+m_{N^{\ast}}^2}{m_{N^{\ast}}^2}M^2.
\end{eqnarray}
We also obtain $u_0$ in terms of the masses of the positive and negative parity baryons as
\begin{eqnarray}
u_0&=&\frac{m_{N^{\ast}}^2}{m_N^2+m_{N^{\ast}}^2},
\end{eqnarray}
at which we calculate the photon wave functions and the related integrals.

As it is clear from Eq. (\ref{sumrules}), to extract the values of the form factors $f_1^T$ and
$f_2^T$ we need to know the residues of both the negative and positive parity nucleons. They are calculated using a two-point correlation function, viz.
\begin{eqnarray}\label{CorrelationFunc1}
\Pi^{\prime}=i \int d^4x~ e^{ip\cdot x}~{\langle}0| {\cal
T}\left ( \eta(x)~\bar{\eta}(0)~\right)|0{\rangle}.
\end{eqnarray}
This  function also is calculated via two different ways, namely
 via hadronic and OPE languages. The hadronic side of this correlation function including both the negative and positive parity nucleons is obtained as
\begin{eqnarray}\label{HadSideNstar}
\Pi^{\prime HAD}=\frac{{\langle}0|\eta|N(p,s){\rangle}
{\langle}N(p,s)|\bar{\eta}|0{\rangle}}{m_N^2-p^{2}}+
\frac{{\langle}0|\eta|N^{\ast}(p,s){\rangle}
{\langle}N^{\ast}(p,s)|\bar{\eta}|0{\rangle}}{m_{N^{\ast}}^2-p^{2}}+....
\end{eqnarray}
 By using the definitions of the residues from Eq. (\ref{matriselements}) and
summing over the spins of the baryons $N$ and $N^{\ast}$ and
applying the Borel transformation, we obtain the  final form of the
hadronic side of the correlation function in the Borel scheme  as
\begin{eqnarray}\label{HadSideN1}
\widehat{\textbf{B}}\Pi^{\prime HAD}=|\lambda_{N}^2|e^{-m_{N}^2/M^{\prime^2}}\Big(\not\!p+m_N U\Big)+
|\lambda_{N^{\ast}}^2|e^{-m_{N^{\ast}}^2/M^{\prime^2}}\Big(-\not\!p+m_{N^{\ast}}U\Big),
\end{eqnarray}
 where $U$ is the unit matrix and $M^{\prime^2}$ is the Borel mass parameters in the mass sum rule.
 The OPE side of the above correlation function is also calculated in deep Euclidean region in terms of QCD degrees of freedom and the structures $\not\!p$ and $U$, i.e.
\begin{eqnarray}
 \widehat{\textbf{B}}\Pi^{\prime OPE}=\Pi^{\prime OPE}_1 (M^{\prime^2},s'_0,\beta)\not\!p+\Pi^{\prime OPE}_2(M^{\prime^2},s'_0,\beta) U,
\end{eqnarray}
where $s'_0$ is the continuum threshold in the mass sum rule.
  Then we match
again the coefficients of the selected structures from both sides. By simultaneous solving of the two appeared equations, we get the following sum rules for the residues of $N$ and
$N^{\ast}$ nucleons in terms of their masses  and other parameters:
\begin{eqnarray}\label{res}
 |\lambda_{N}^2|=\frac{e^{m_{N}^2/M^{\prime^2}}}{m_N+m_{N^{\ast}}}\Big[m_{N^{\ast}}\Pi^{\prime OPE}_1 (M^{\prime^2},s'_0,\beta)+\Pi^{\prime OPE}_2(M^{\prime^2},s'_0,\beta)\Big],\nonumber\\
|\lambda_{N^{\ast}}^2|=\frac{e^{m_{N^{\ast}}^2/M^{\prime^2}}}{m_N+m_{N^{\ast}}}\Big[-m_{N}\Pi^{\prime OPE}_1 (M^{\prime^2},s'_0,\beta)+\Pi^{\prime OPE}_2(M^{\prime^2},s'_0,\beta)\Big].
\end{eqnarray}
Here we should also stress that the functions $\Pi^{\prime OPE}_1 (M^{\prime^2},s'_0,\beta)$ and $\Pi^{\prime OPE}_2 (M^{\prime^2},s'_0,\beta)$ have very lengthy expressions and for this reason
 we do not present their
explicit forms.

\section{Numerical results}
First of all we would like to obtain the behaviors of the residues of the positive and negative parity baryons with respect to the general parameter $\beta$ entering the general form of the interpolating current discussed
in the previous section. To this aim we use some input parameters $\uu = \dd= -(0.243)^3~GeV^3$, $\sp = 0.8
\uu$ and $m_0^2 = (0.8\pm0.2)~GeV^2$ \cite{Belyaev}. Beside these, we need also to find the working regions of auxiliary parameters $M^{\prime^2}$, $s'_0$ and $\beta$ entering the  sum rules for the residues.
The general criteria in the sum rule method is that the physical quantities are practically independent of the auxiliary parameters, hence, we look for some regions of these helping parameters such that the residues
weakly depend of them in these regions. The continuum threshold depends on the energy of the first excited state with the same quantum numbers as the interpolating current for both the negative and positive parity states.
Our calculations show that in the intervals
\begin{eqnarray}
1.08\,GeV^2\leq &s_0^{\prime}&\leq 1.98 \,GeV^2 \,\,\,\,\,\mbox{for}
\,\,\,\,\, N,
\nonumber \\
\mbox{and}
\nonumber \\
2.51 \,GeV^2\leq &s_0^{\prime}&\leq 2.73\,GeV^2 \,\,\,\,\, \mbox{for}
\,\,\,\,\, N^{\ast},
\end{eqnarray}
the residues weakly depend on the continuum threshold. The working region of the Borel mass parameter  $M^{\prime^2}$ is obtained demanding that not only the perturbative contributions
to the residues exceed the non-perturbative ones but also the operators with higher dimensions have small contributions compared to the lower dimensional operators. This leads to the Borel windows
\begin{eqnarray}
0.5 \,GeV^2\leq &M^{\prime^2}&\leq 1.5\,GeV^2 \,\,\,\,\,\mbox{for}
\,\,\,\,\, N,
\nonumber \\
\mbox{and}
\nonumber \\
2\,GeV^2\leq &M^{\prime^2}&\leq 4\,GeV^2 \,\,\,\,\, \mbox{for} \,\,\,\,\,
N^{\ast}.
\end{eqnarray}
It is also found that in the intervals
$-0.60\leq \cos \theta  \leq -0.30$ and $ 0.30 \leq \cos \theta \leq 0.60 $ for $N$ as well as  $-0.62 \leq \cos \theta  \leq -0.30 $ for
$ N^{\ast}$, at which $\theta=tan^{-1}\beta$, the residues have relatively weak dependency on the general parameter $\beta$.
Using these working windows for the  auxiliary parameters as well as other inputs we find that the residue squares as functions of the general parameter $\beta$
in the positive and negative parity channels can be parametrized
 as
\begin{eqnarray}
\lambda_N^2(\beta)&=&e^{\frac{m_N^2}{M^{\prime^2}}}\Phi_1(\beta),
\nonumber \\
\lambda_{N^{\ast}}^2(\beta)&=&e^{\frac{m_{N^{\ast}}^2}{M^{\prime^2}}}
\Phi_2(\beta),
\end{eqnarray}
where
\begin{eqnarray}
\Phi_1(\beta)&=&[2.28-3.20\beta+64.10\beta^{2}]\times10^{-6} ~~(GeV^6),
\nonumber \\
\Phi_2(\beta)&=&[-7.38 -4.20 \beta + 2.82\beta^2]\times10^{-5}~~(GeV^6).
\end{eqnarray}

Now, we perform numerical analysis of the sum rules
for the transition form factors and use them to predict the decay width and
branching ratio of the radiative  $N^{\ast}\rightarrow N\gamma$ transition. To
this aim, we need the photon DAs and wave functions as well as the corresponding parameters, which are  given in the Appendix.
Beside these inputs and the residues obtained above in terms of the general parameter $\beta$, we also need the parameters 
$f_{3 \gamma} = - 0.0039~GeV^2$ \cite{Ball} and  $\chi(1~GeV)=-4.4~GeV^{-2}$ \cite{Kogan}.
%

The sum rules   for the form  factors $f_1^T$ and $f_2^T$, contain three more auxiliary parameters:  the Borel mass
parameter $M^2$, the continuum threshold $s_{0}$ and the arbitrary
parameter $\beta$ entering  the expression of the interpolating
current. The working window for $M^2$ is
obtained requiring that not only  the ground state contribution exceeds the contributions of the higher
states and continuum
but  the contributions of the leading twists exceed those of the higher twists, i.e., the series of the sum rules for the form factors converge.  These requirements lead to the interval $2~GeV^2\leq
M^{2}\leq4~GeV^2 $. We also choose the interval $2.51~GeV^2\leq s_0\leq2.73~GeV^2$ for the continuum threshold, at which the dependence of the form factors on this parameter are weak.
Our numerical
calculations also lead to the working regions $-0.60\leq \cos\theta
\leq -0.30$ and $0.30\leq \cos\theta \leq 0.60$ for the arbitrary parameter $\beta=\tan\theta$.
We depict the dependence of the form factors $f_1^T$ and $f_2^T$  on
the Borel mass parameters $M^2$ at fixed value of $s_0$ and different values of $\cos\theta$ in figure 1. With a quick glance
at this figure, we see that these form factors depict weak
dependence on the Borel parameter on its working regions.
\begin{figure}[h!]
\includegraphics[width=8cm]{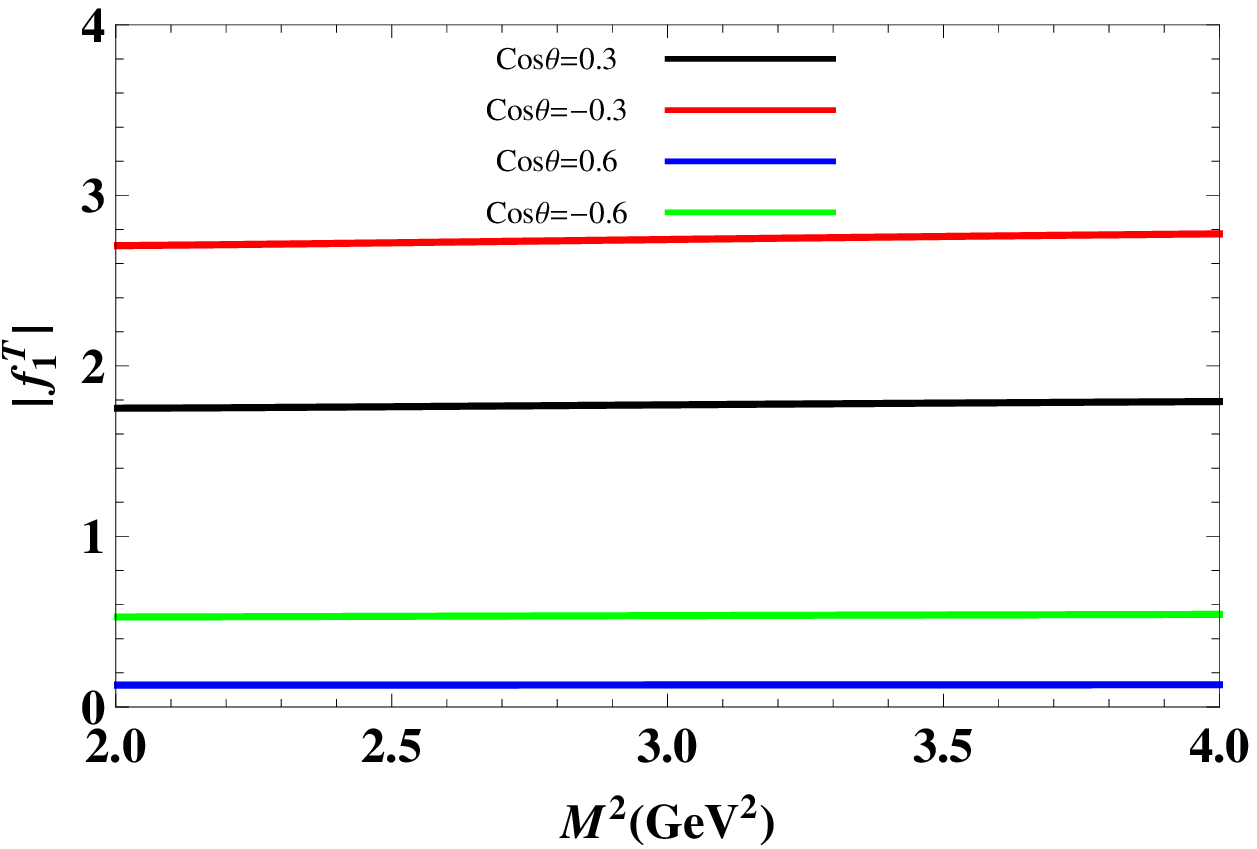}
\includegraphics[width=8cm]{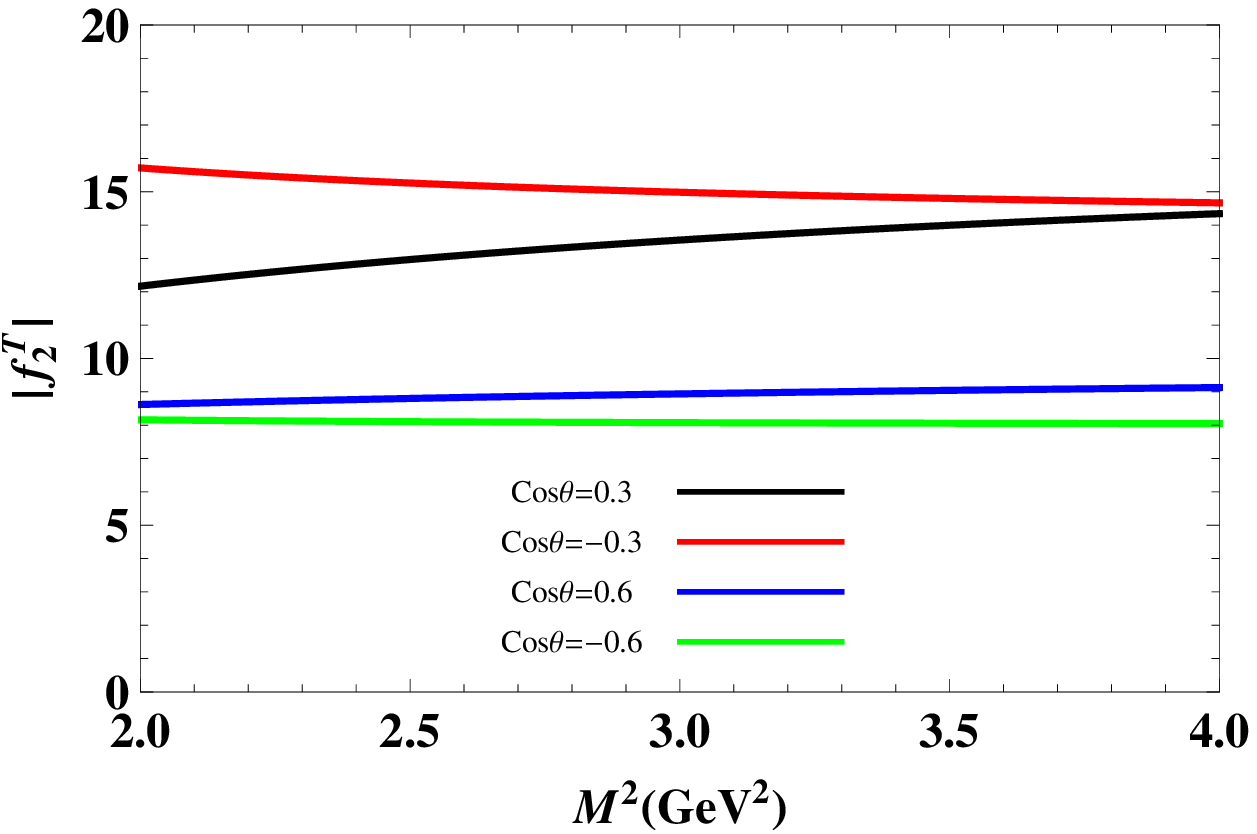}
\caption{ \textbf{Left:} The dependence of the form factor $f_1^T$
for $N^{\ast}\rightarrow N\gamma$ transition on the Borel mass
parameter $M^{2}$ at $s_0=2.67~GeV^2$ and different values of $\cos\theta$. \textbf{Right:} The dependence of
the form factor $f_2^T$ for $N^{\ast}\rightarrow N\gamma$
transition on the Borel mass parameter $M^{2}$ at $s_0=2.67~GeV^2$ and different values of $\cos\theta$. }
\label{fig1}
\end{figure}
 Using all input parameters and working regions of all auxiliary parameters as well as the masses and residues of the negative and positive parity baryons obtained via mass sum rules, the average  values of the
 transition form factors $f_1^T$ and $f_2^T$ for the
$N^{\ast}\rightarrow N\gamma$ channel are obtained as presented in table 1.

\begin{table}[h]
\renewcommand{\arraystretch}{1.5}
\addtolength{\arraycolsep}{3pt}
$$
\begin{array}{|c|c|}
\hline \hline
       \mbox{$f_1^T$}    & 1.26 \pm0.37     \\
\hline
  \mbox{$f_2^T$} &11.18\pm3.35 \\
                          \hline \hline
\end{array}
$$
\caption{The  values of the form factors $f_1^T$ and
$f_2^T$  for the $N^{\ast}\rightarrow
N\gamma$ transition.} \label{formfctors}
\renewcommand{\arraystretch}{1}
\addtolength{\arraycolsep}{-1.0pt}
\end{table}

At the end of this section, we would like to calculate the decay
width for  the radiative $N^{\ast}\rightarrow N\gamma$ transition.
Considering the corresponding transition matrix
elements in Eq. (\ref{matriselements}), we obtain the following
formula for the total width of the  transition under consideration:

\begin{eqnarray}
\Gamma&=&\frac{\alpha_{em}}{2m_{N^{\ast}}^3}\Big(m_N^2-m_{N^{\ast}}^2\Big)\Bigg\{|f_1^{T}|^2
\Big(m_N^2+4m_Nm_{N^{\ast}}+m_{N^{\ast}}^2\Big)-3Re(f_1^Tf_2^T)\Big(m_{N^{\ast}}^2-m_N^2\Big)
\nonumber \\
&+&|f_2^T|^2\Big(m_{N^{\ast}}-m_N\Big)^2\Bigg\}.
\end{eqnarray}

Using the numerical values for the form factors and other input parameters together with the width of the $N^*$ state  \cite{PDG}, we obtain the numerical  values
for the decay width and branching ratio of the radiative $N^{\ast}\rightarrow
N\gamma$ transition as presented in table 2. In comparison, we also depict the existing related experimental data in this table. Looking at this table we see that our prediction for the
branching fraction of the considered transition is in a very good consistency with the experimental data.
\begin{table}[h]
\renewcommand{\arraystretch}{1.5}
\addtolength{\arraycolsep}{3pt}
$$
\begin{array}{|c|c|c|}
\hline \hline
       \mbox{}    & \Gamma(GeV)& Br    \\
\hline
  \mbox{Present work} &(3.08\pm0.09)\times 10^{-4}& (0.21\pm0.06)\times 10^{-2}
   \\
   \hline
  \mbox{PDG\cite{PDG}} &-& (0.15-0.30)\%
   \\
                          \hline \hline
\end{array}
$$
\caption{Decay width and branching ratio of the radiative  $N^{\ast}\rightarrow
N\gamma$ transition.} \label{formfctors}
\renewcommand{\arraystretch}{1}
\addtolength{\arraycolsep}{-1.0pt}
\end{table}

In summary, we have calculated the transition form factors  responsible for the radiative  transition of negative to positive parity nucleon in the frame work of the LCSR using the most general for of the interpolating
current coupling to both the negative and positive parity nucleons as well as the photon DAs. We found the working regions of all auxiliary parameters and obtained the behavior of the residues in terms of the general mixing
 parameter entering the interpolating current. We used them to predict the numerical values of the form factors for the real photon. The values of the transition form factors are then used to
estimate the decay width and branching fraction of the transition under consideration. Our result on the branching ratio of the radiative  $N^{\ast}\rightarrow
N\gamma$ transition is in a good agreement with the existing experimental data.

\section{Acknowledgment}
This work has been supported in part by the Scientific and Technological
Research Council of Turkey (TUBITAK) under the research project 114F018.

\section*{Appendix }

The matrix elements appearing in our calculations  parametrized in terms of  the photon DAs are written as (see \cite{Ball}) :

 \begin{eqnarray}
&&\langle \gamma(q) \vert  \bar q(x) \sigma_{\mu \nu} q(0) \vert 0
\rangle  = -i e_q \bar q q (\varepsilon_\mu q_\nu -
\varepsilon_\nu q_\mu) \int_0^1 du e^{i \bar u qx} \left(\chi
\varphi_\gamma(u) + \frac{x^2}{16} \mathbb{A}  (u) \right)
\nonumber \\ && -\frac{i}{2(qx)}  e_q \qq \left[x_\nu
\left(\varepsilon_\mu - q_\mu \frac{\varepsilon x}{qx}\right) -
x_\mu \left(\varepsilon_\nu - q_\nu \frac{\varepsilon x}{q
x}\right) \right] \int_0^1 du e^{i \bar u q x} h_\gamma(u),
\nonumber \\
&&\langle \gamma(q) \vert  \bar q(x) \gamma_\mu q(0) \vert 0
\rangle = e_q f_{3 \gamma} \left(\varepsilon_\mu - q_\mu
\frac{\varepsilon x}{q x} \right) \int_0^1 du e^{i \bar u q x}
\psi^v(u),
\nonumber \\
&&\langle \gamma(q) \vert \bar q(x) \gamma_\mu \gamma_5 q(0) \vert
0 \rangle  = - \frac{1}{4} e_q f_{3 \gamma} \epsilon_{\mu \nu
\alpha \beta } \varepsilon^\nu q^\alpha x^\beta \int_0^1 du e^{i
\bar u q x} \psi^a(u),
\nonumber \\
&&\langle \gamma(q) | \bar q(x) g_s G_{\mu \nu} (v x) q(0) \vert 0
\rangle = -i e_q \qq \left(\varepsilon_\mu q_\nu - \varepsilon_\nu
q_\mu \right) \int {\cal D}\alpha_i e^{i (\alpha_{\bar q} + v
\alpha_g) q x} {\cal S}(\alpha_i),
\nonumber \\
&&\langle \gamma(q) | \bar q(x) g_s \tilde G_{\mu \nu} i \gamma_5
(v x) q(0) \vert 0 \rangle = -i e_q \qq \left(\varepsilon_\mu
q_\nu - \varepsilon_\nu q_\mu \right) \int {\cal D}\alpha_i e^{i
(\alpha_{\bar q} + v \alpha_g) q x} \tilde {\cal S}(\alpha_i),
\nonumber \\
&&\langle \gamma(q) \vert \bar q(x) g_s \tilde G_{\mu \nu}(v x)
\gamma_\alpha \gamma_5 q(0) \vert 0 \rangle = e_q f_{3 \gamma}
q_\alpha (\varepsilon_\mu q_\nu - \varepsilon_\nu q_\mu) \int
{\cal D}\alpha_i e^{i (\alpha_{\bar q} + v \alpha_g) q x} {\cal
A}(\alpha_i),
\nonumber \\
&&\langle \gamma(q) \vert \bar q(x) g_s G_{\mu \nu}(v x) i
\gamma_\alpha q(0) \vert 0 \rangle = e_q f_{3 \gamma} q_\alpha
(\varepsilon_\mu q_\nu - \varepsilon_\nu q_\mu) \int {\cal
D}\alpha_i e^{i (\alpha_{\bar q} + v \alpha_g) q x} {\cal
V}(\alpha_i) ,\nonumber \\ && \langle \gamma(q) \vert \bar q(x)
\sigma_{\alpha \beta} g_s G_{\mu \nu}(v x) q(0) \vert 0 \rangle  =
e_q \qq \left\{
        \left[\left(\varepsilon_\mu - q_\mu \frac{\varepsilon x}{q x}\right)\left(g_{\alpha \nu} -
        \frac{1}{qx} (q_\alpha x_\nu + q_\nu x_\alpha)\right) \right. \right. q_\beta
\nonumber \\ && -
         \left(\varepsilon_\mu - q_\mu \frac{\varepsilon x}{q x}\right)\left(g_{\beta \nu} -
        \frac{1}{qx} (q_\beta x_\nu + q_\nu x_\beta)\right) q_\alpha
\nonumber \\ && -
         \left(\varepsilon_\nu - q_\nu \frac{\varepsilon x}{q x}\right)\left(g_{\alpha \mu} -
        \frac{1}{qx} (q_\alpha x_\mu + q_\mu x_\alpha)\right) q_\beta
\nonumber \\ &&+
         \left. \left(\varepsilon_\nu - q_\nu \frac{\varepsilon x}{q.x}\right)\left( g_{\beta \mu} -
        \frac{1}{qx} (q_\beta x_\mu + q_\mu x_\beta)\right) q_\alpha \right]
   \int {\cal D}\alpha_i e^{i (\alpha_{\bar q} + v \alpha_g) qx} {\cal T}_1(\alpha_i)
\nonumber \\ &&+
        \left[\left(\varepsilon_\alpha - q_\alpha \frac{\varepsilon x}{qx}\right)
        \left(g_{\mu \beta} - \frac{1}{qx}(q_\mu x_\beta + q_\beta x_\mu)\right) \right. q_\nu
\nonumber \\ &&-
         \left(\varepsilon_\alpha - q_\alpha \frac{\varepsilon x}{qx}\right)
        \left(g_{\nu \beta} - \frac{1}{qx}(q_\nu x_\beta + q_\beta x_\nu)\right)  q_\mu
\nonumber \\ && -
         \left(\varepsilon_\beta - q_\beta \frac{\varepsilon x}{qx}\right)
        \left(g_{\mu \alpha} - \frac{1}{qx}(q_\mu x_\alpha + q_\alpha x_\mu)\right) q_\nu
\nonumber \\ &&+
         \left. \left(\varepsilon_\beta - q_\beta \frac{\varepsilon x}{qx}\right)
        \left(g_{\nu \alpha} - \frac{1}{qx}(q_\nu x_\alpha + q_\alpha x_\nu) \right) q_\mu
        \right]
    \int {\cal D} \alpha_i e^{i (\alpha_{\bar q} + v \alpha_g) qx} {\cal T}_2(\alpha_i)
\nonumber \\ &&+
        \frac{1}{qx} (q_\mu x_\nu - q_\nu x_\mu)
        (\varepsilon_\alpha q_\beta - \varepsilon_\beta q_\alpha)
    \int {\cal D} \alpha_i e^{i (\alpha_{\bar q} + v \alpha_g) qx} {\cal T}_3(\alpha_i)
\nonumber \\ &&+
        \left. \frac{1}{qx} (q_\alpha x_\beta - q_\beta x_\alpha)
        (\varepsilon_\mu q_\nu - \varepsilon_\nu q_\mu)
    \int {\cal D} \alpha_i e^{i (\alpha_{\bar q} + v \alpha_g) qx} {\cal T}_4(\alpha_i)
                        \right\},
\end{eqnarray}
where $\varphi_\gamma(u)$ is the leading twist-2 photon DAs, the photon DAs' $\psi^v(u)$,
$\psi^a(u)$, ${\cal A}(\alpha_i)$ and ${\cal V}(\alpha_i)$ have twist 3; and
$h_\gamma(u)$, $\mathbb{A}$(u) and ${\cal T}_i$ ($i=1,~2,~3,~4$) have twist 4  \cite{Ball}. In the above relations  $\chi$ is the magnetic
susceptibility of the light quarks.

The measure $\int{\cal D} \alpha_i$ entering the calculations  is defined as
\begin{equation}
\int {\cal D} \alpha_i = \int_0^1 d \alpha_{\bar q} \int_0^1 d
\alpha_q \int_0^1 d \alpha_g \delta(1-\alpha_{\bar
q}-\alpha_q-\alpha_g).\nonumber \\
\end{equation}
The above mentioned  photon DAs are also parametrized as \cite{Ball}:
\begin{eqnarray}
\varphi_\gamma(u) &=& 6 u \bar u \left( 1 + \varphi_2(\mu)
C_2^{\frac{3}{2}}(u - \bar u) \right),
\nonumber \\
\psi^v(u) &=& 3 \left(3 (2 u - 1)^2 -1 \right)+\frac{3}{64}
\left(15 w^V_\gamma - 5 w^A_\gamma\right)
                        \left(3 - 30 (2 u - 1)^2 + 35 (2 u -1)^4
                        \right),
\nonumber \\
\psi^a(u) &=& \left(1- (2 u -1)^2\right)\left(5 (2 u -1)^2
-1\right) \frac{5}{2}
    \left(1 + \frac{9}{16} w^V_\gamma - \frac{3}{16} w^A_\gamma
    \right),
\nonumber \\
{\cal A}(\alpha_i) &=& 360 \alpha_q \alpha_{\bar q} \alpha_g^2
        \left(1 + w^A_\gamma \frac{1}{2} (7 \alpha_g - 3)\right),
\nonumber \\
{\cal V}(\alpha_i) &=& 540 w^V_\gamma (\alpha_q - \alpha_{\bar q})
\alpha_q \alpha_{\bar q}
                \alpha_g^2,
\nonumber \\
h_\gamma(u) &=& - 10 \left(1 + 2 \kappa^+\right)
C_2^{\frac{1}{2}}(u - \bar u),
\nonumber \\
\mathbb{A}(u) &=& 40 u^2 \bar u^2 \left(3 \kappa - \kappa^+
+1\right) \nonumber \\ && +
        8 (\zeta_2^+ - 3 \zeta_2) \left[u \bar u (2 + 13 u \bar u) \right.
\nonumber \\ && + \left.
                2 u^3 (10 -15 u + 6 u^2) \ln(u) + 2 \bar u^3 (10 - 15 \bar u + 6 \bar u^2)
        \ln(\bar u) \right],
\nonumber \\
{\cal T}_1(\alpha_i) &=& -120 (3 \zeta_2 + \zeta_2^+)(\alpha_{\bar
q} - \alpha_q)
        \alpha_{\bar q} \alpha_q \alpha_g,
\nonumber \\
{\cal T}_2(\alpha_i) &=& 30 \alpha_g^2 (\alpha_{\bar q} -
\alpha_q)
    \left((\kappa - \kappa^+) + (\zeta_1 - \zeta_1^+)(1 - 2\alpha_g) +
    \zeta_2 (3 - 4 \alpha_g)\right),
\nonumber \\
{\cal T}_3(\alpha_i) &=& - 120 (3 \zeta_2 -
\zeta_2^+)(\alpha_{\bar q} -\alpha_q)
        \alpha_{\bar q} \alpha_q \alpha_g,
\nonumber \\
{\cal T}_4(\alpha_i) &=& 30 \alpha_g^2 (\alpha_{\bar q} -
\alpha_q)
    \left((\kappa + \kappa^+) + (\zeta_1 + \zeta_1^+)(1 - 2\alpha_g) +
    \zeta_2 (3 - 4 \alpha_g)\right),\nonumber \\
{\cal S}(\alpha_i) &=& 30\alpha_g^2\{(\kappa +
\kappa^+)(1-\alpha_g)+(\zeta_1 + \zeta_1^+)(1 - \alpha_g)(1 -
2\alpha_g)\nonumber \\&+&\zeta_2
[3 (\alpha_{\bar q} - \alpha_q)^2-\alpha_g(1 - \alpha_g)]\},\nonumber \\
\tilde {\cal S}(\alpha_i) &=&-30\alpha_g^2\{(\kappa -
\kappa^+)(1-\alpha_g)+(\zeta_1 - \zeta_1^+)(1 - \alpha_g)(1 -
2\alpha_g)\nonumber \\&+&\zeta_2 [3 (\alpha_{\bar q} -
\alpha_q)^2-\alpha_g(1 - \alpha_g)]\},
\end{eqnarray}
where, the values of the different  parameters  inside the wave functions are given as
 $\varphi_2(1~GeV) = 0$, $w^V_\gamma = 3.8 \pm 1.8$,
$w^A_\gamma = -2.1 \pm 1.0$, $\kappa = 0.2$, $\kappa^+ = 0$,
$\zeta_1 = 0.4$, $\zeta_2 = 0.3$, $\zeta_1^+ = 0$ and $\zeta_2^+ =
0$ \cite{Ball}.


\begin{thebibliography}{99}

\bibitem{Jefferson} V. Punjabi et al., Phys. Rev. C 71, 055202 (2005).
\bibitem{Mainz1} B. Krusche, and S. Schadmand, Prog. Part. Nucl. Phys. 51, 399 (2003).
\bibitem{Mainz2} M. Kortulla et al., Phys. Rev. Lett. 89, 272001 (2002).
\bibitem{Mainz3} M. Kortulla et al., Phys. Rev. Lett. 61, 147 (2008).

\bibitem{Chung} Y. Chung, H. G. Dosch, M. Kremer and D. Schall, Nucl. Phys. B1 97, 55 (1982).
\bibitem{Jido} D. Jido , N. Kodama and M. Oka,  Phys. Rev. D 54, 4532 (1996).
\bibitem{Oka} M. Oka, D. Jido, A. Hosaka, Nucl. Phys. A 629, 156 (1998).
\bibitem{Lee} Frank X. Lee, Derek B. Leinweber,  Nucl. Phys. Proc. Suppl.73, 258 (1999).
\bibitem{Kondo} Y. Kondo, O. Morimatsu,  T. Nishikawa,  Nucl. Phys. A 764, 303 (2006).


\bibitem{V.M. Braun} V. M. Braun et al., Phys. Rev. D 89, 094511 (2014).

\bibitem{Aliev}  T. M. Aliev, M. Savci, Phys. Rev. D 89, 053003 (2014).
\bibitem{Narodetskii} I. M. Narodetskii, M.A. Trusov, JETP Letters, 99,  57 (2014).

\bibitem{savci1} T. M. Aliev, M. Savci, Phys. Rev. D 88, 056021 (2013).
\bibitem{savci2} T. M. Aliev, M. Savci, Phys. Rev. D 90, 096012 (2014).
\bibitem{savci3} T. M. Aliev, M. Savci, J. Phys. G 41, 075007  (2014).

\bibitem{Aliev1}  T. M. Aliev, A. Ozpineci, M. Savci, Phys. Rev. D 66, 016002, (2002); Erratum-ibid.D67, 039901,
(2003).
\bibitem{Balitsky} I. I.  Balitsky,  V. M.  Braun,  Nucl. Phys.  B  311, 541 (1989).
\bibitem{Braun2} V. M.  Braun, I. E. Filyanov, Z. Phys. C 48, 239 (1990).
\bibitem{Belyaev} V. M. Belyaev,  B. L.  Ioffe, JETP  56, 493 (1982).
\bibitem{Ball} P.  Ball,  V. M.  Braun, N. Kivel,  Nucl. Phys.  B  649, 263 (2003).
\bibitem{Kogan} V. M. Belyaev,  I. I.  Kogan, Yad. Fiz.  40, 1035 (1984).
\bibitem{PDG} K.A. Olive et al. (Particle Data Group), Chin. Phys. C, 38, 090001 (2014).









\end{thebibliography}
\end{document}